\documentstyle[trans]{rspublic}
\def\bi{\begin{itemize}}
\def\ei{\end{itemize}}
\def\bq{\begin{quotation}}
\def\eq{\end{quotation}}

\def\thedemobiblio#1{\smallskip\par
 \list{}{\labelwidth 0pt \leftmargin 1em \itemindent -1em \itemsep 1pt}
 \small \parindent 0pt
 \parskip 1.5pt plus .1pt\relax
 \def\newblock{\hskip .11em plus .33em minus .07em}
 \sloppy\clubpenalty4000\widowpenalty4000
 \sfcode`\.=1000\relax}

\begin{document}

\title[Creation of maximally entangled pairs of distant atoms]{Quantum communication and 
the creation of maximally entangled pairs of atoms over a noisy channel}

\author[H.-J. Briegel, W. D\"ur, S. J. van Enk, J. I. Cirac, and P. Zoller]
{H.-J. Briegel$^{1,2}$,
W. D\"ur$^1$, S. J. van Enk$^3$, J. I. Cirac$^{1,2}$, \\ and P. Zoller$^1$}

\affiliation{$^1$ Institut f{\"u}r Theoretische Physik,
Universit{\"a}t Innsbruck, Technikerstra{\ss}e 25, A-6020 Innsbruck, Austria\\
$^2$ Universidad de Castilla--La Mancha, 13071 Ciudad Real, Spain\\
$^3$ Norman Bridge Laboratory of Physics 12-33, California Institute of Technology, 
Pasadena CA 91125, USA}

\maketitle

\label{firstpage}
\begin{abstract}
We show how to create maximally entangled EPR pairs
between spatially distant atoms, each of them inside a high-Q optical
cavity, by sending photons through a general, noisy channel, such as a
standard optical fiber.
An error correction scheme that uses few auxiliary atoms in each cavity
effectively eliminates photoabsorption and other transmission errors.
This realizes the `absorption free channel.'
A concatenation protocol using the absorption free channel allows for 
quantum communication with single qubits over distances much larger than 
the coherence length of the channel.    

\end{abstract}

\section{Introduction}

The concept of a distant maximally entangled pair of particles, i.e.\ an 
EPR pair (Einstein {\em et al.} 1935) plays a key role in both
theoretical and in practical aspects of quantum communication.
It has been shown by Bennett {\em et. al} (1993) that EPR pairs can be used 
to send unknown quantum states by means of teleportation. Other examples include
the possibility of secure distribution of cryptographic keys (Ekert 1991),
imperfect cloning of a quantum state (Gisin \& Massar 1997, Bru{\ss} {\em et al.} 1997), 
and the implementation of nonlocal conditional operations in quantum networks (Grover 1997).

In the present paper we concentrate on a realistic scheme that
could be used for the creation of EPR pairs between atoms, and 
thus for quantum communication. We use a quantum optical implementation 
that combines elements that have been experimentally realized or can be
expected to be realized in the near future.
For example, significant process has been reported in experimental
quantum optics with trapped ions and in cavity QED, both in the optical and
in the microwave regime (Nobel Symposium 1997).
Highlights of this development include first steps towards
the implementation of {\em quantum-logical operations}, such as two-qubit
operations that form the building blocks for quantum computing.
Monroe {\em et al.} (1995) have succeeded in performing a universal 2-bit 
quantum gate, using the quantized center-of-mass motion of a trapped ion 
and its internal
quantum degrees of freedom. In  CQED  the basic elements of gate
operations involving the internal state of an atom and a cavity mode have
been demonstrated by Turchette {\em et al.} (1995), and recently  entangled internal
states of two atoms  have been produced (Hagley {\em et al.} 1997).

This paper gives a brief introduction to recent work  
of the Innsbruck group (Cirac {\em et al.} 1997a,b; van Enk {\em et al.} 
1997a,b,c) on the subject of quantum communication.

\section{Physical scheme}

\begin{figure}[ht]
\begin{center}
\begin{picture}(200,70)
\put(0,0){\epsfxsize=200pt\epsffile[126 354 513 467]{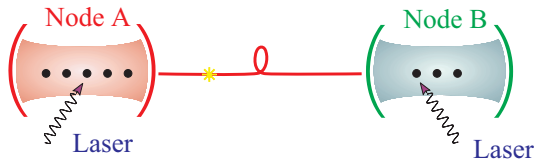}}
\end{picture}
\end{center}
\caption[]{Quantum optical scheme for creating an EPR pair of atoms over a
noisy channel.}
\label{FLBczmk}
\end{figure}

The scheme we propose is decribed in detail by Cirac {\em et al.} (1997a) and 
van Enk {\em et al.} (1997a).
It uses longlived states of atoms as the physical basis for storing qubits,
and photons as a means for transfering these qubits from one atom to another.
To allow for a controlled transfer of the qubit, the atoms are embedded in high
finesse optical cavities which are connected by an optical fiber, as shown in 
Fig.~\ref{FLBczmk}. For
successful transmission of photons, appropriately tailored Raman pulses are 
applied to the atoms at the sending and the receiving time. These pulses
map the qubit from the atomic state to a specific photon wavepacket and
{\em vice versa}.

\input{epsf}

The optical cavity-fiber system, together with the laser
pulses constitutes what we abstractly call a {\em noisy quantum channel}.
Before we talk about noise processes in more detail, let us briefly discuss the 
ideal situation.

Ideally, the scheme realizes the following transmission
\begin{equation}
  \big[\chi_0|0\rangle_A+\chi_1|1\rangle_A\big]|0\rangle_B
  \longrightarrow
  |0\rangle_A\big[\chi_0|0\rangle_B+\chi_1|1\rangle_B\big]
\label{swap}
\end{equation}
where an unknown superposition of internal states $|0\rangle$
and $|1\rangle$ in atom $A$ in the first cavity 
is transferred to atom $B$ in the second 
cavity. The cavities may be part of a larger 
network, so we often refer to them as node A and node B, respectively.
The selected internal states $|0\rangle$ and $|1\rangle$
of the atoms define, in the language of quantum information theory, the
`computational basis' for the qubit.

It is important to realize that the atom $A$ may be entangled to other atoms
in the same cavity or at other nodes of the network. In that situation,
the coefficients $\chi_0$ and $\chi_1$ in (\ref{swap}) are no longer complex
numbers but denote unnormalized states of the other atoms.
Thus the transmission (\ref{swap}) can be used to transfer single atomic
states, but also to transfer {\em entanglement}.
For instance, starting from single particle states, an EPR pair can be
created by a two-step process
\begin{eqnarray}
 \big[\chi_0|0\rangle_A+\chi_1|1\rangle_A\big]|0\rangle_{A_2}|0\rangle_B
 &\longrightarrow &
 \big[\chi_0|0\rangle_{A_2}|0\rangle_A
 + \chi_1|1\rangle_{A_2}|1\rangle_A\big]|0\rangle_B\nonumber\\
 &\longrightarrow &
 |0\rangle_{A_2}\big[\chi_0|0\rangle_A|0\rangle_B
 + \chi_1|1\rangle_A|1\rangle_B\big]\,.
\label{ekert}
\end{eqnarray}
Here, the first arrow refers to a {\em local} CNOT operation
between two atoms $A$ and $A_2$ in the first cavity.
The second arrow transfers the state of $A_2$ to $B$, thereby transfering
the entanglement between the atoms $A$ and $A_2$ to an entanglement between
atoms $A$ and $B$.
At the end of this composite transformation, the state of the
auxiliary atom $A_2$ is the same as initially and {\em factors out}.
For $\chi_0=\chi_1$, an ideal EPR pair is created.

From this simple consideration we conclude: The creation of
a distant EPR pair is possible if
1.\ the local CNOT operation and 2.\ the transfer operation (\ref{swap})
can be realized without errors, or if these errors can be corrected.
Corresponding to these operations, two different kinds of errors
need to be corrected: 1. local (gate) errors, 2. transfer errors.

When the photons are sent along optical fibers, 
then photoabsorption will be a dominant transfer error.
Losses may also occur by incoherent scattering on the surface of the
cavity mirrors and at the coupling segments between the cavities and the
fiber. Another typical transfer error will be caused by imperfectly designed 
laser pulses for the Raman transition. An example for a local gate error
is spontaneous emission in one of the atoms during the gate operation.

In this paper, we will assume that local gate 
operations can be performed without errors, and concentrate on
errors that occur during the photon transfer. A discussion
of error correction for 2-bit gate operations can be found 
in the paper by van Enk {\em et al.} (1997c).

\section{The photonic channel}

In a realistic model, we have to consider the possibility that
the transfer of the atomic state from cavity A to B  is imperfect.
There is a certain probability that the atom in B will not be 
excited, even though A was excited. This is due
to the interaction of the compound atom-cavity-fiber system with the
environment which, even if small, in principle always exists.
This results in an entanglement of the atomic states in (\ref{swap})
with the environment, i.e. the cavity walls, the fiber, and the radiation
field of the free space.

In the following, we assume that photons can be absorbed but not created
by the channel. 
This is a very good approximation for optical photons, where the mean thermal
number of photons in the cavities and the fiber is exceedingly small.
In this situation, the most general expression for an imperfect transfer
operation is of the form
\begin{eqnarray}
 |0\rangle_A |0\rangle_B |E\rangle &\longrightarrow &
 |0\rangle_A |0\rangle_B |E_0\rangle
 \nonumber \\
 |1\rangle_A |0\rangle_B |E\rangle &\longrightarrow &
 |0\rangle_A |1\rangle_B |E_1\rangle
 + |0\rangle_A |0\rangle_B |E_a\rangle
\label{swapE}
\end{eqnarray}
where $|E\rangle,|E_0\rangle, \dots $ denote unnormalized states of the
environment. It is expedient to write 
$|E_0\rangle = {\cal T}_0 |E\rangle$,
$|E_1\rangle = {\cal T}_1 |E\rangle$,
$|E_a\rangle = {\cal T}_a |E\rangle$,
thereby introducing operators that entangle the system with the environment.
With this notation, (\ref{swapE}) can be expressed in the compact 
form\footnote{In expressions of this type, 
it is understood that both the left and the right-hand side is applied on a 
given state of the environment. Using this compact notation keeps the expressions 
much more transparent when twofold or more complex applications of the channel 
are studied.}
\begin{eqnarray}
 |0\rangle_A |0\rangle_B &\longrightarrow &
 |0\rangle_A |0\rangle_B {\cal T}_0
 \nonumber \\
 |1\rangle_A |0\rangle_B &\longrightarrow &
 |0\rangle_A |1\rangle_B {\cal T}_1
 + |0\rangle_A |0\rangle_B {\cal T}_a
\label{photonic_channel}
\end{eqnarray}
which defines the photonic channel (van Enk {\em et al.} 1997b).

The optical cavities together with the fiber form a compound optical
system with a certain resonant structure that defines its spectrum of
quasi modes, its relaxation constants, etc.
In the special case when only photoabsorption plays a role, the
operators in (\ref{photonic_channel}) have a simple form.
For optical frequencies, the state of the environment can be 
very well approximated by the vacuum state, so one can write 
$ {\cal T}_0 = 1$,
$ {\cal T}_1 = \alpha(\tau) \sim e^{-\kappa\tau}$,
$ {\cal T}_a = \sum_j \beta_j(\tau) b_j^{\dag}$,
with
$\sum_j |\beta_j(\tau)|^2 \sim 1 - e^{-2\kappa \tau}$
where $\kappa$ is the damping rate of the total (atom-)cavity-fiber
system, and $\tau$ is the transfer time.
The operators $b_j^{\dag}$, $b_j$ are amplitude operators of the $j$ th
oscillator mode of the environment.

More generally, the operators ${\cal T}_{0,1,a}$ in (\ref{photonic_channel})
may describe spontaneous emission processes, photon absorption, as well as 
transitions to and repumping from other internal states of the atoms. 
Thus, all complicated physics is hidden in the three operators.
In this general (non-stationary) situation, the time dependence of the
environmental terms has to be taken into account.
The operators ${\cal T}_{0,1,a}$ then depend
on the initial time when the transfer starts.
As a consequence, when iterating the channel (\ref{photonic_channel}),
the temporal ordering of the operators becomes important, e.g.
${\cal T}_1(t_1){\cal T}_0(t_0) \ne {\cal T}_0(t_1){\cal T}_1(t_0)$.

When using (\ref{photonic_channel}) to create an EPR pair as in
(\ref{ekert}),
we obtain
\begin{eqnarray}
 \big[\chi_0|0\rangle_A + \chi_1|1\rangle_A\big]|0\rangle_B
 &\longrightarrow &
 \big[\chi_0|0\rangle_A|0\rangle_B{\cal T}_0
 + \chi_1|1\rangle_A|1\rangle_B{\cal T}_1\big] \nonumber\\
 && + \chi_1|1\rangle_A|0\rangle_B{\cal T}_a\,.
\end{eqnarray}

For $\chi_0=\chi_1$, this expression can be written in the form
\footnote{Throughout this paper, normalization factors are omitted unless
they are needed.}
\begin{equation}
|\Phi^{+}_{AB}\rangle \big[{\cal T}_0+{\cal T}_1\big]
      + |\Phi^{-}_{AB}\rangle \big[{\cal T}_0-{\cal T}_1\big]
      + \left( |\Psi^{+}_{AB}\rangle+|\Psi^{-}_{AB}\rangle \right){\cal T}_a
\label{ekertT}
\end{equation}
where we use the Bell basis
$|\Phi^{\pm}_{AB}\rangle$$= \frac{1}{\sqrt{2}}
 \left( |0\rangle_A|0\rangle_B\pm|1\rangle_A|1\rangle_B \right)$ and
$ |\Psi^{\pm}_{AB}\rangle = \frac{1}{\sqrt{2}}
 \left( |0\rangle_A|1\rangle_B\pm|1\rangle_A|0\rangle_B \right)$.
The fidelity of the resulting pair (\ref{ekertT}) can be defined by its overlap 
with the ideal result $|\Phi^{+}_{AB}\rangle$. This overlap is given by the norm
\begin{equation}
 F=\left\|\frac{[{\cal T}_0+{\cal T}_1]|E\rangle}{2}\right\|^2 
 \sim \left|\frac{1+e^{-\kappa\tau}}{2}\right|^2\,.
\end{equation}
The estimate of $F$ in the second term demonstrates how the coupling of the modes of the
cavity-fiber system to the environment reduces the attainable fidelity
of the EPR pair.
In particular, $F$ decreases exponentially with the transfer time and the corresponding
length of the fiber.

In order to create an EPR pair over a distance comparable or larger than
the absorption length of the photonic channel, we need to find a method to
{\em detect and correct} a photon loss that may occur during the transfer.
Loosely speaking, we are seeking to eliminate the absorption term
${\cal T}_a$ in
(\ref{ekertT}), and to minimize the other term ${\cal T}_0-{\cal T}_1$.

In the following, we outline a method that uses one resp.\ two auxiliary 
atoms in each cavity. This outline just summarizes the essential steps. 
For details, the reader should consult the papers by van Enk {\em et al.} (1997a,b)
and Cirac {\em et al.} (1997b).

\section{Error correction and purification with finite means}\label{SECafc}

The main idea is to entangle the atom in the first cavity with auxiliary 
(backup) atoms, before transmitting the information. This is
reminiscent of a redundant coding scheme, with the fundamental difference
that our scheme allows to correct errors to {\em all orders} in the
photoabsorption probability.
By measuring a certain joint state of two atoms in the receiver cavity,
one is able to {\em detect} a photon loss while {\em maintaining} the
initial coherence of the atomic state that was sent. Therefore, the
transmission can be repeated as often as necessary until no error is
detected.

In detail, this requires three steps:\newline
(1) {\em Encoding} of the atomic state into a three-particle entangled state
\begin{eqnarray}
 \chi_0|0\rangle_A+\chi_1|1\rangle_A &\longrightarrow &
 \chi_0\big[|0\rangle_A|0\rangle_{A_2}|0\rangle_{A_3}
          + |1\rangle_A|1\rangle_{A_2}|1\rangle_{A_3} \big] \nonumber\\
 && +  \chi_1\big[|0\rangle_A|0\rangle_{A_2}|1\rangle_{A_3}
          + |1\rangle_A|1\rangle_{A_2}|0\rangle_{A_3} \big]\,.
\label{encoding}
\end{eqnarray}
This can be realized by applying two CNOT operations between $A_3$ and $A$,
and $A$ and $A_2$, respectively.\newline
(2) {\em Transmission} of a photon {\em twice} by using (\ref{photonic_channel})
between atom $A_2$ and $B_2$ and then between $A_2$ and $B$, applying
a local flip operation on $A$ in between. The result of this operation
is a multiparticle entangled state (Cirac {\em et al.} 1997b) whose explicit 
form will not be given here.\newline
(3) {\em Measuring} the states of certain backup atoms in both cavities.
Combined with appropriate local unitary transformations,
one obtains one of two results. 

The effect of this procedure is summarized in the following {\em absorption-free 
channel}
\begin{eqnarray}
 \big[\chi_0|0\rangle_A+\chi_1|1\rangle_A\big]|0\rangle_B
 &\longrightarrow &
 \chi_0|0\rangle_A|0\rangle_B{\cal S}_0
 +\chi_1|1\rangle_A|1\rangle_B{\cal S}_1 \nonumber\\
 && \nonumber\\
 && \big[\chi_0|0\rangle_A+\chi_1|1\rangle_A\big]|0\rangle_B
    {\cal S}_a\,.
\label{AFC}
\end{eqnarray}
\begin{picture}(0,0)
\put(160,42){\vector(1,-1){18}}
\put(147,22){\footnotesize\tt error}
\end{picture}
Owing to the twofold transmission process, the operators $\cal S$
appearing in (\ref{AFC}) are products of the $\cal T$ operators, e.g.\
${\cal S}_0={\cal T}_0{\cal T}_1$, ${\cal S}_1={\cal T}_1{\cal T}_0$, or
in different order.
The important feature to notice is that, depending on the results of the
measurement in step (3), two outcomes are possible: If an error is detected, 
the state is projected onto the second line of (\ref{AFC}) and the 
transmission can be repeated; if no error is detected, the state is 
projected onto the first line of (\ref{AFC}), which completes the channel.

By using (\ref{AFC}) instead of (\ref{photonic_channel}) one obtains
\begin{eqnarray}
 \big[|0\rangle_A + |1\rangle_A\big]|0\rangle_B
 &\longrightarrow &
 |0\rangle_A|0\rangle_B{\cal S}_0
 + |1\rangle_A|1\rangle_B{\cal S}_1 
 \nonumber\\
 && = |\Phi^{+}_{AB}\rangle \frac{1}{2}\big[{\cal S}_0+{\cal S}_1\big]
      + |\Phi^{-}_{AB}\rangle \frac{1}{2}\big[{\cal S}_0-{\cal S}_1\big]\,.
\label{ekertS}
\end{eqnarray}

For the simple example considered after (\ref{photonic_channel}), with 
${\cal T}_0=1$ and ${\cal T}_1=e^{-\kappa\tau}$, we have 
${\cal S}_0=e^{-\kappa\tau}$ and ${\cal S}_1=e^{-\kappa\tau}$, thus
the second term in (\ref{ekertS}) vanishes. In this situation,
an ideal EPR pair is established after a {\em single} use of the channel (\ref{AFC}). This
corresponds to an average number of phototransmissions of 
$e^{2\kappa\tau}$.  

More generally, a similar result is obtained when the state of the environment 
does not depend on the temporal ordering of the operators ${\cal T}_0$ and ${\cal T}_1$.
Such a {\em stationary environment} is defined by ${\cal T}_1(t_1){\cal T}_0(t_0)|E\rangle
={\cal T}_0(t_1){\cal T}_1(t_0)|E\rangle$, i.e.\ ${\cal S}_0|E\rangle={\cal S}_1|E\rangle$.
For any system with a stationary environment, an ideal EPR pair is created by a single 
application of (\ref{AFC}).

For the discussion of the general, non-stationary case, let us first rewrite 
the result (\ref{ekertS}) in the form
\begin{equation}
 |\Psi^{(1)}\rangle = |\Phi^{+}_{AB}\rangle |E_{+}^{(1)}\rangle
                    + |\Phi^{-}_{AB}\rangle |E_{-}^{(1)}\rangle
\label{ekertSE}
\end{equation}
where $|E_{\pm}^{(1)}\rangle=\frac{1}{2}({\cal S}_0 \pm {\cal
S}_1)|E\rangle$. The norm (square) of the environment $|E_{+}^{(1)}\rangle$
determines the fidelity of the pair.

\begin{figure}[ht]
\begin{center}
\begin{picture}(400,250)
\put(0,10){\epsfxsize=160pt\epsffile[76 64 540 726]{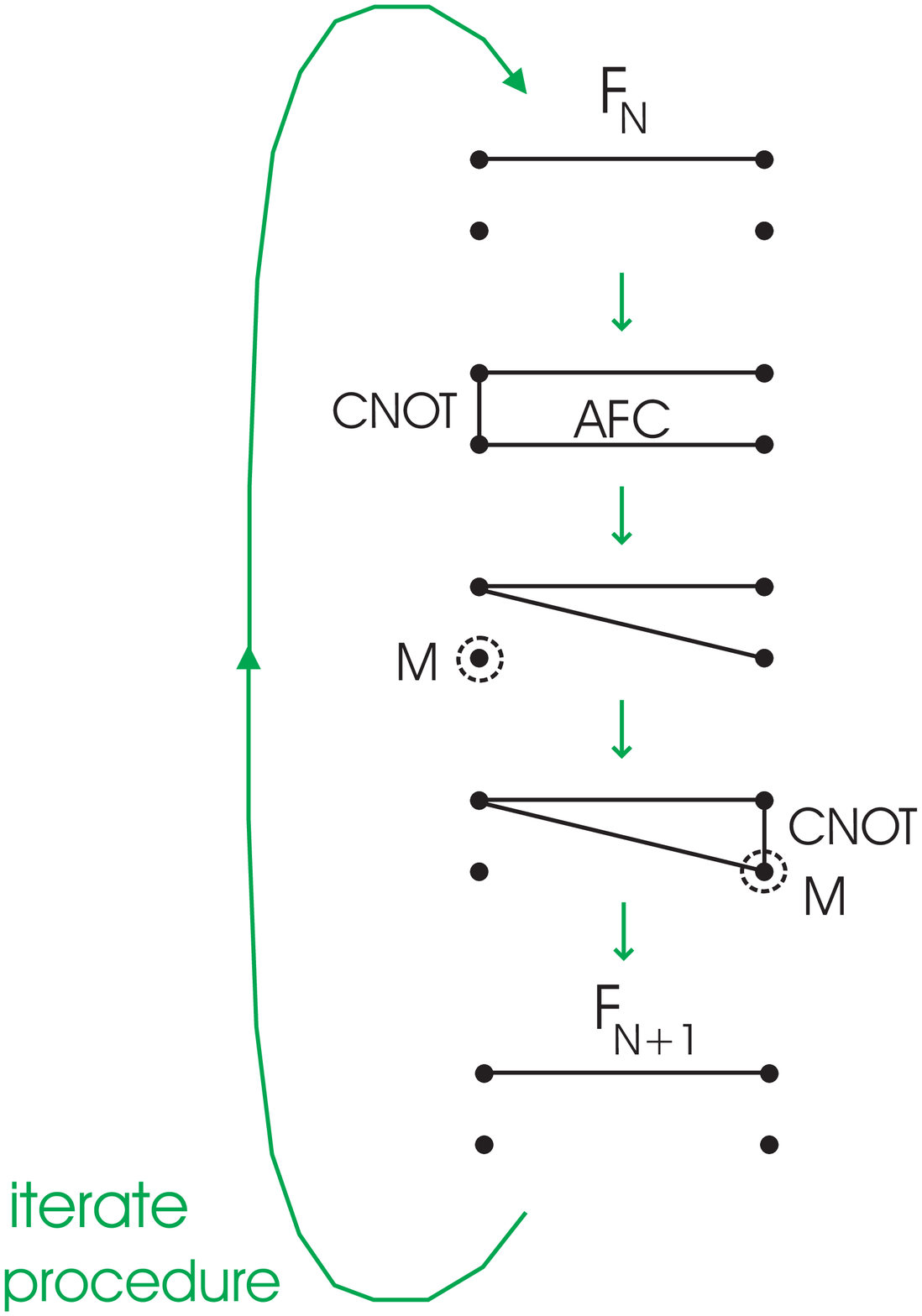}}
\put(200,20){\epsfxsize=180pt\epsffile[77 238 523 609]{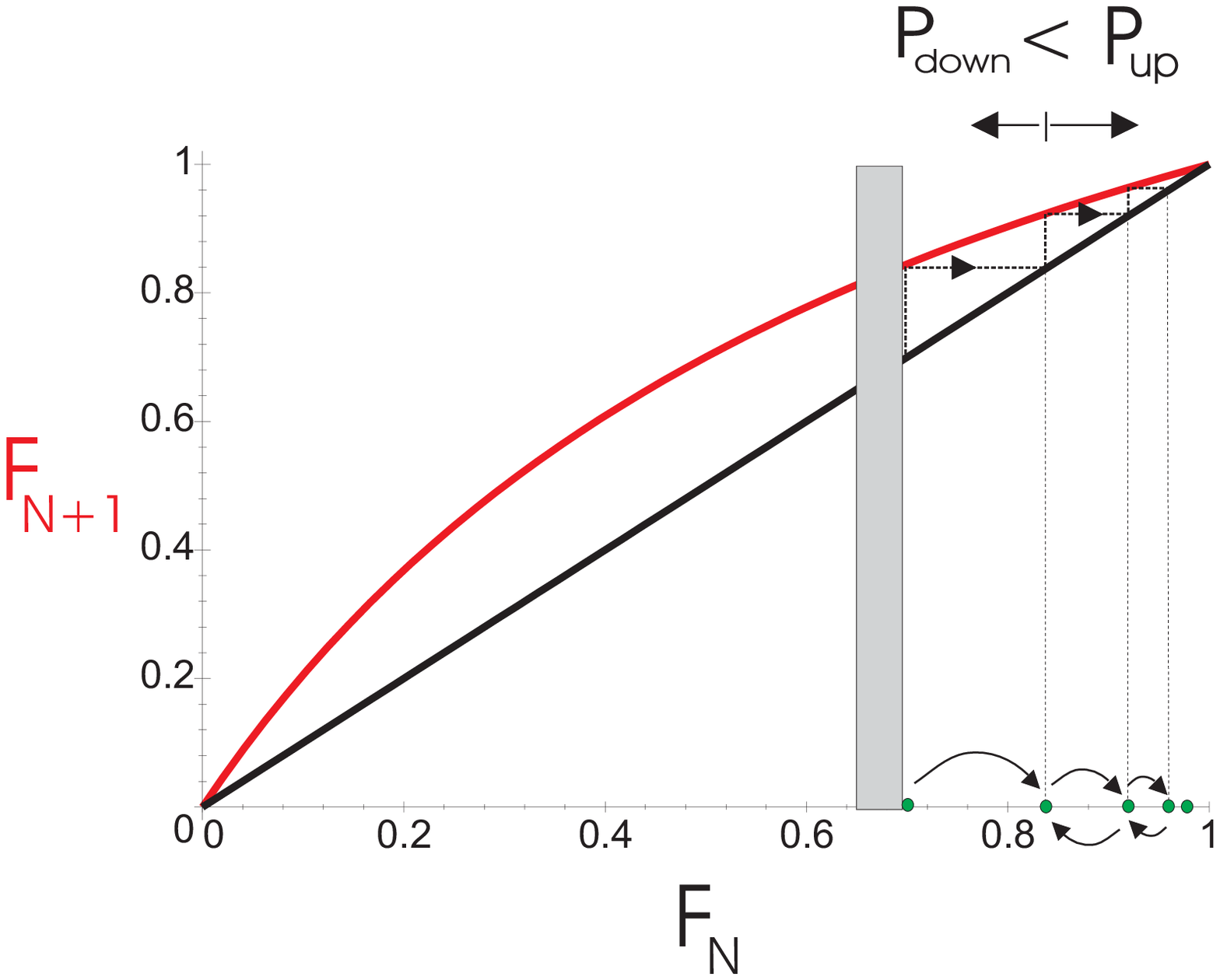}}
\put(0,200){(a)}
\put(200,200){(b)}
\end{picture}
\end{center}
\caption[]{Purification of an EPR pair with finite means.
(a) Iterative purification protocol. At each purification step, an EPR pair of the 
form (\ref{ekertSEN}) with fidelity $F_N$ is temporarily entangled with two auxiliary
atoms. This involves two CNOT operations, the absorption-free channel AFC, and 
measurements M. Furthermore, there are some Hadamard transformations that are not 
shown in the figure. The value of the new fidelity $F_{N+1}$ depends on the result of 
the measurements M. Note that this scheme operates on the {\em same 
set of atoms} at each step, thereby realizing a `self-purification process.'    
(b) One-sided random-walk process for fidelity.
After each iteration step in (a), the fidelity $F_N$ increases (decreases) with a 
certain probability $P_{\rm up}$ ($P_{\rm down}$)) that depends on $N$. 
If $F_N$ happens to drop below the initial value $F_0$, we reset the pair to this value
by a single use of the AFC, as in (\ref{ekertS}). This is equivalent to a one-sided 
random walk process with reflections at a lower barrier at $F_0$, as indicated in the 
figure. On average, the fidelity thereby approaches unity 
exponentially fast, $F_N \sim 1-e^{-{\rm const.}\times N}$.}
\label{FLBmunch}
\end{figure}

At this point, the key advantage of the absorption-free channel (AFC) comes
into play, namely that it corrects errors in the transmission process while
maintaining the coherence and possible entanglement of the state it is 
applied to. This allows for an iterative purification protocol 
(van Enk {\em et al.} 1997b).
At each purification step, the pair is temporarily entangled with two auxiliary
atoms, one at each node, using both local CNOT operations and the AFC.
In some sense, this creates an auxiliary EPR pair
that is used to purify (\ref{ekertSE}). 
The detailed protocol is shown schematically in Fig.~\ref{FLBmunch}a.

\input{epsf}

This protocol transforms (\ref{ekertSE}) into a sequence of states of the form
\begin{equation}
 |\Psi^{(N)}\rangle = |\Phi^{+}_{AB}\rangle |E_{+}^{(N)}\rangle
                    + |\Phi^{-}_{AB}\rangle |E_{-}^{(N)}\rangle
\label{ekertSEN}
\end{equation}
where either

$|E_{\pm}^{(N)}\rangle = \frac{1}{2}({\cal S}_0 \pm {\cal
S}_1)|E_{\pm}^{(N-1)}\rangle $
or
$ |E_{\pm}^{(N)}\rangle = \frac{1}{2}({\cal S}_0 \mp {\cal
S}_1)|E_{\pm}^{(N-1)}\rangle$,
depending on the result of the measurement.
In the first case, which happens with probability $P_{\rm up}=P_{\rm up}^{(N)}$, 
the fidelity of the pair increases. In the second case, which happens with
$P_{\rm down}=1-P_{\rm up}$, the fidelity decreases. 
One can show that this creates a stochastic process corresponding to a
one-sided random-walk process as depicted in Fig.~\ref{FLBmunch}b.
On average, the fidelity $F_N=\langle E_{+}^{(N)}|E_{+}^{(N)} \rangle$
thereby converges towards unity exponentially fast with the number of
purification steps.

It is quite remarkable that this describes a finite system that iteratively 
purifies itself.
We note that standard purification protocols (Bennett {\em et al.} 1996, 
Deutsch {\em et al.} 1996, Gisin 1996), 
assume a large number of EPR pairs which need to be coherently 
manipulated (connected). In contrast, the present scheme uses iterated 
operations on {\em few} atoms.

\section{Communication over long distances}

With the methods discussed in the previous sections, it is possible to create 
an EPR pair of high fidelity by sending single photons through a dissipative 
and noisy channel that connects the atoms.

There is, however, a limitation to the method when the transmission 
time through the channel becomes much larger than its relaxation time, i.e.\ 
if $\kappa\tau \gg 1$. As the absorption probability grows exponentially with 
$\tau$, so will the required number of  repetitions for one  successful transmission.
We have to keep in mind that in the implementation of the AFC (\ref{AFC}), we 
have assumed that all local operations are error free. In reality, these operations 
will never be perfect, but there will be  small probabilities for failure. 
As the number of required transmissions increases, the error probabilites
will eventually add up and spoil the coherence-preserving action of the AFC.

Absorption losses are well-known in problems of electric signal transmission
through classical channels where, at regularly spaced intervals,
amplifiers are put in the channel. In classical (digital) communication technique
such amplifiers are not only used to amplify but also to restore the signal. 
The distance between the amplifiers is then determined by (1) the damping rate of the fiber
and (2) the bit rate of the transmission (dispersion effects). In the classical case, 
dispersion effects are more important whereas in the quantum case, when we send {\em single 
photons}, damping will be the main problem. 

In order build up the EPR correlations, single qubits (photons) need to be transmitted 
which cannot be amplified (Wootters \& Zurek 1982, Glauber 1986).  
All we can do here is to detect whether a photon has been absorbed and, whenever 
that is the case, repeat the transmission. 

\begin{figure}[ht]	
\begin{center}
\begin{picture}(300,130)
\put(0,80){\epsfxsize=300pt\epsffile[0 0 652 23]{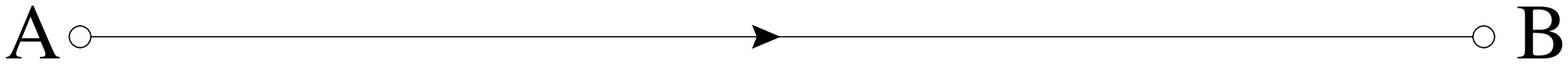}}
\put(0,30){\epsfxsize=300pt\epsffile[0 0 656 23]{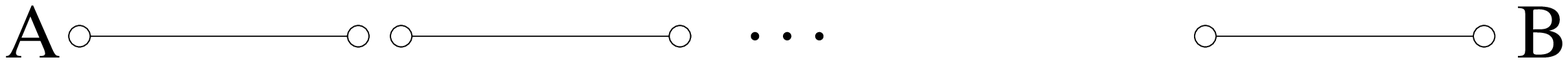}}
\put(140,97){$e^{l/l_0}$}\put(140,70){$l$}
\put(90,47){$e^{l/Nl_0}$}\put(90,20){$l/N$}
\end{picture}
\end{center}
\caption[]{Simple and compound fiber for transmission of single photons from $A$ to $B$.
As with classical amplifiers, to transmit single photons reliable over long distances, 
the fiber has to be divided into several segments, at the end of which transmission 
errors are measured.}
\label{FLBfiber}
\end{figure}

For any practical implementation of the scheme, it is crucial to know how the the 
number of operations scales with the length of the channel, i.e. the distance between 
the atoms. For the following discussion, let us assume that the dominant transmission error 
is given by photoabsorption. 
This corresponds to a photonic channel (\ref{photonic_channel}) with
${\cal T}_0=1$ and ${\cal T}_1=e^{-\kappa \tau}=e^{-l/2l_0}$ where $l_0=c/2\kappa$ 
defines the half length of the fiber. 
The probability for a successful transmission
from $A$ to $B$, as indicated in Fig.~\ref{FLBfiber}a, is then $p(l)=e^{-l/l_0}$ 
where $l$ is the length of the fiber. 
Correspondingly, the average number of required repetitions is  
\begin{equation}
 n(l) = \frac{1}{p(l)} = e^{l/l_0}\,. 
\label{simple_fiber}
\end{equation}
It is clear that this leads to unrealisticly high numbers for any experiment,
if the fiber is much longer than a few half lengths $l_0$.

\input{epsf}

In Fig.~\ref{FLBfiber}b, this situation is compared with a compound fiber, consisting 
of a sequence of $N$ segments of length $l/N$ each.
At the end of every segment, there is a `checkpoint' at which it is measured
whether a transmission error has occurred. This can be done with the method 
explained in Sec.~\ref{SECafc} by using a few ions in a cavity.
If an absorption error is detected, the transmission across that segment is
repeated. Then a photon is sent through the subsequent segment, and so on.
The average total number of repetitions on each segment is $n(l/N)=e^{l/l_0N}$.
Correspondingly, the total number of transmissions required for successfully
sending the photon across the {\em compound fiber} is 
\begin{equation}
 n_{\mbox{\scriptsize com}} = \frac{N}{p(l/N)} = Ne^{l/Nl_0}\,. 
\label{n_com}
\end{equation}
This is to be compared with (\ref{simple_fiber}). 
The compound fiber is preferable to the simple fiber if 
\begin{equation}
 Ne^{l/Nl_0} < e^{l/l_0}\,.
\end{equation}
The optimum number of segments is given by the value of $N$ that minimizes the
left-hand side of above equation, which is $N_{\mbox{\scriptsize min}} = l/l_0$.

The minimum number of transmissions along the compound fiber is thus
given by (\ref{n_com}) with $N=N_{\mbox{\scriptsize min}}$, that is
\begin{equation}
 n_{\mbox{\scriptsize min}}=N_{\mbox{\scriptsize min}}e^{l/N_{\mbox{\scriptsize min}}l_0} = l/l_0 e^1\,.
\label{compound_transmission}
\end{equation}
This optimum situation is realized if the checkpoints are placed along the fiber 
with a spacing corresponding to the half length $l_0$. 

An example: The fiber used in the recent cryptographic experiments by the Geneva 
group (Zbinden {\em et al.} 1996, Tittel {\em et al.} 1997) 
has a half length $l_0$ of the order of 10 km. Using such a fiber
for establishing an EPR pair between Innsbruck and Oxford, say, 
($l\sim 1000$km$=100 l_0$) requires $100 e=272$ transmissions if one uses 
a compound fiber with 100 checkpoints. For a simple fiber, with no checkpoint, this average 
number is as high as $e^{100}=2.69\times 10^{43}$! 
For 1 transmission per microsecond, the total time needed in the first case would be
of the order of 1 millisecond; in the latter case it would be larger than the age of the universe.

\medskip
So far we have only talked about `successfully sending a photon across the fiber.'
What we really have in mind, however, is not to directly send arbitrary states of 
photons, but to establish an EPR pair between distant atoms $A$ and $B$ with the 
aid of this fiber. Although these two problems are closely related, there is an
important difference. To build up the EPR correlations, we need only to send 
very specific states of photons, namely, according to (\ref{ekertS}), those which 
correspond to the atomic state $(|0\rangle+|1\rangle)/\sqrt{2}$. This has the 
advantage that we do not need to trace the evolution of some arbitrary qubit 
throughout the channel, including the checkpoints, simply because there is no 
information processing. The channel may be used for quantum communication {\em after} 
we have established the EPR correlations between $A$ and $B$ but, at that stage,
the  state of the channel is irrelevant.\footnote{The fiber need not even exist any more.}

The problem we are dealing with can be rephrased in more general terms, without refering to 
a particular physical realization of the channel. For this, we divide the problem into two steps.
To create an EPR pair over a noisy channel of length $l$, which  is composed of $N$ segments 
of length $l/N$, we  (1)  establish $N$ (short-distance)  EPR pairs, one pair across every segment;
(2) connect the EPR pairs by making Bell measurements at each checkpoint, 
followed by classical communication between the checkpoints. 
As a result,  a (long-distance) EPR pair is created between the uttermost atoms. 

\begin{figure}[ht]
\begin{center}
\begin{picture}(300,100)
\put(0,20){\epsfxsize=300pt\epsffile[0 0 607 88]{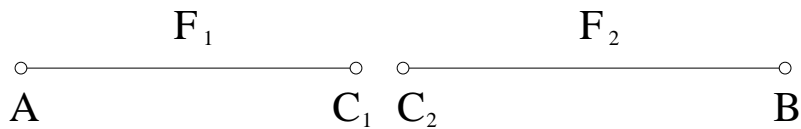}}
\end{picture}
\end{center}
\caption[]{Connection of two EPR pairs $A\leftrightarrow C_1$ and $C_2\leftrightarrow B$ 
with fidelity $F_1$ and $F_2$, resp., at checkpoint $C$. 
This creates a new pair $A\leftrightarrow B$  between $A$ and $B$ 
with fidelity $F \le \mbox{\scriptsize min}\{F_1,F_2\}$.}
\label{FLBconnect}
\end{figure}

For a compound fiber consisting of just two segments,
this scheme is shown in Fig.~\ref{FLBconnect}. The connection process, i.e.\ step (2),  
is equivalent to a teleportation of the state of particle $C_1$,  which is entangled with particle  $A$, 
to $B$.

To study the connection process in more detail, let us go back and relax our previous assumption 
that only absorption errors occur, while maintaining  the assumption that local 2 bit 
operations can be performed without error. For example, suppose the timing of the 
Raman pulses at the beginning and at the end of the transmission is not perfect.
In this case, on each segment  an EPR pair 
of the form (\ref{ekertSEN}) is created as the result of an $N$ step process, with a fidelity 
$F_N=\langle E_{+}^{(N)}|E_{+}^{(N)} \rangle$ that may be high but is not equal to 
unity.\footnote{The actual value of $F_N$ will be determined by the maximum number of purification 
steps one is willing to apply, or by the limitation inherent in the implementation 
of the AFC, due to imperfections of the local operations in step (1) that leads to  
(\ref{encoding}).} 

When connecting two EPR pairs of the form (\ref{ekertSEN}), with fidelity 
$F_1=F_{N_1}$ and $F_2=F_{N_2}$, respectively, as depicted in Fig.~\ref{FLBconnect}, the resulting pair
has a fidelity that is smaller than $F_1$ and $F_2$. 
A brief calculation shows that the fidelity $F$ of the new pair satisfies 
\begin{equation}
 2F-1 = (2F_1-1)(2F_2-1)\,.
\label{connect}
\end{equation}
Upon connecting $N$ pairs of fidelity $F_0$, we obtain one distant pair of fidelity
\begin{equation}
 F = \frac{1}{2}\left\{ 1+(2F_0-1)^N \right\}
\label{F0N}
\end{equation}
which can be seen by iterating (\ref{connect}) $N$ times. Note that the connecting
need not be done sequentially, i.e. first at checkpoint one, then at two, and 
so on. A faster (or rather the fastest) way to do this would be in step 1 to 
simultaneously connect the pairs at every second checkpoint, which leaves one 
with $N/2$ pairs of fidelity $F_1=\frac{1}{2}\left\{ 1+(2F_0-1)^2 \right\}$.
In step 2 the remaining pairs are again connected at every second checkpoint,
which results in $N/4$ pairs with fidelity $F_2=\frac{1}{2}\left\{ 1+(2F_0-1)^4 \right\}$. If 
$N=2^n$, then the final EPR pair with fidelity (\ref{F0N}) is obtain by
iteration after a total of $n=\mbox{log}_2N$ steps.  

The lesson to be learned from (\ref{F0N}) is the following. 
The fidelity decreases exponentially with the number of segments
unless the initial fidelity $F_0$ on every segment is unity.
By dividing the fiber into shorter segments, we have thus eliminated the effect 
of exponentially increasing transmission losses while, at the same time,  
introducing an exponential decrease in the fidelity.

The drawback of the second effect, however, does not completely ruin the 
gain from the first effect.  Since the purification 
protocol described by the sequence (\ref{ekertSEN}) converges 
towards unit fidelity exponentially fast, one can
compensate for the exponential decrease of $F$   
in (\ref{F0N})  by starting from a higher initial value $F_0$. 
This requires only a few extra purification steps on each segment, 
before they are connected.  There is a limit to this, which is given 
by the maximum fidelity  attainable on each segment. Furthermore, the 
connection process itself will in general be imperfect. Even if we started from 
$N$ EPR pairs of unity fidelity, any imperfection in the connection 
operation would decrease the fidelity of the connected pair, and therefore
the fidelity of the connection operation would enter exponentially 
and reduce the final fidelity.  

In summary, any imperfections in local operations (i.e.\ 2 bit, 1 bit and  
measurements)  result in a fidelity that decreases exponentially 
with the length of the channel. 
Conversely, the number of operations required to 
produce an EPR pair of a given fidelity increases exponentially 
with the distance of the particles.

This poses a severe limitation to any distant quantum communication.

We just report here that there is a way around this exponential  
scaling property by using a purification protocol which re-purifies 
the obtained EPR pairs after a number of connecting steps 
(Cirac {\em et al.} unpublished).
The idea of the procedure is similar to what has been called 
a `concatenated quantum code' by Knill \& Laflamme (1996) in the 
context of fault tolerant quantum computing. \newline
However, the problem of creating an EPR pair via a  
noisy channel -consisting of a sequence of imperfect gate 
operations and transfer operations-  is different from 
fault tolerant quantum computation.  
The crucial difference is that we are not, in the first place,  sending an 
arbitrary qubit through the channel. 
In other words, {\em while creating the EPR correlations, there is no quantum 
information being processed},  although the EPR pair may subsequently be used 
for communication via teleportation.

As a result, we obtain fidelity requirements on the local operations  
which are in the few-percent region. This should be compared with the  
number of $10^{-5}$ (Knill \& Laflamme 1996) for fault tolerant quantum 
computing.

\section{Conclusion}

We have outlined a procedure for creating maximally entangled pairs 
of atoms which are connected by a noisy quantum channel.  
The essential features of this procedure are (1) the usage of 
{\em finite means} i.e. a few ions in two resonators; (2) the implementation of an 
{\em absorption-free channel} that is used for non-local entanglement 
operations between ions in different resonators; and 
(3) a {\em concatenated purification} algorithm that allows for 
communication beyond a certain distance that, in usual approaches, is 
imposed the by exponential transmission losses.

\begin{acknowledgments}
This work was supported in part by the \"Osterreichischer Fonds zur F\"orderung
der wis\-sen\-schaft\-lich\-en Forschung, and by the European TMR network 
ERB-FMRX-CT96-0087. 
\end{acknowledgments}


\label{lastpage}

\end{document}